\documentclass[aps,prl,twocolumn,showpacs,floatfix,superscriptaddress]{revtex4}
\usepackage{graphicx}
\begin{document}
\flushbottom
\title{Direct observation of stress accumulation and relaxation in small bundles of superconducting vortices in tungsten thin-films}
\author{I. Guillam\'on \footnote{Present adress:H.H. Wills Physics Laboratory, University of Bristol, Tyndall Avenue, Bristol BS8 1TL, UK}}
\affiliation{Laboratorio de Bajas Temperaturas, Departamento de
F\'isica de la Materia Condensada, Instituto de Ciencia de
Materiales Nicol\'as Cabrera, Facultad de Ciencias \\ Universidad
Aut\'onoma de Madrid, E-28049 Madrid, Spain}
\author{H. Suderow}
\affiliation{Laboratorio de Bajas Temperaturas, Departamento de
F\'isica de la Materia Condensada, Instituto de Ciencia de
Materiales Nicol\'as Cabrera, Facultad de Ciencias \\ Universidad
Aut\'onoma de Madrid, E-28049 Madrid, Spain}
\author{S. Vieira}
\affiliation{Laboratorio de Bajas Temperaturas, Departamento de
F\'isica de la Materia Condensada, Instituto de Ciencia de
Materiales Nicol\'as Cabrera, Facultad de Ciencias \\ Universidad
Aut\'onoma de Madrid, E-28049 Madrid, Spain}
\author{J. Ses\'e}
\affiliation{Instituto de Nanociencia de Arag\'on, Universidad de Zaragoza, Zaragoza, 50018, Spain}
\affiliation{Departamento de F\'isica de la Materia Condensada, Universidad de Zaragoza, 50009 Zaragoza, Spain}
\author{R. C\'ordoba}
\affiliation{Instituto de Nanociencia de Arag\'on, Universidad de Zaragoza, Zaragoza, 50018, Spain}
\affiliation{Departamento de F\'isica de la Materia Condensada, Universidad de Zaragoza, 50009 Zaragoza, Spain}
\author{J.M. De Teresa}
\affiliation{Instituto de Ciencia de Materiales de Arag\'on, Universidad de Zaragoza-CSIC, Facultad de Ciencias, Zaragoza, 50009, Spain}
\affiliation{Departamento de F\'isica de la Materia Condensada, Universidad de Zaragoza, 50009 Zaragoza, Spain}
\author{M.R. Ibarra}
\affiliation{Instituto de Nanociencia de Arag\'on, Universidad de Zaragoza, Zaragoza, 50018, Spain}
\affiliation{Instituto de Ciencia de Materiales de Arag\'on, Universidad de Zaragoza-CSIC, Facultad de Ciencias, Zaragoza, 50009, Spain}
\affiliation{Departamento de F\'isica de la Materia Condensada, Universidad de Zaragoza, 50009 Zaragoza, Spain}

\begin{abstract}
We study the behavior of bundles of superconducting vortices when increasing the magnetic field using scanning tunneling microscopy and spectroscopy (STM/S) at 100 mK. Pinning centers are given by features on the surface corrugation. We find strong net vortex motion in a bundle towards a well defined direction. We observe continuos changes of the vortex arrangements, and identify small displacements, which stress and deform the vortex bundle, separated by larger re-arrangements or avalanches, which release accumulated stress.
\end{abstract}
\pacs{74.55.+v, 74.25.Wx, 74.78.-w} \date{\today} \maketitle

Among the most relevant topics in the physics of type II superconductors is the knowledge and control over flux line arrangements. Applications of these materials always require fixing vortices by an adequate distribution of pinning centers.  The characteristics and behavior of both, vortices and pinning centers, are critically dependent on the structural and superconducting properties of each sample. However, there are some important aspects which have general character. Usually, flux enters the sample from the edge in form of Abrikosov vortices, which are driven towards the interior by the Lorentz force due to the Meissner shielding currents. As first proposed by Bean \cite{Bean64}, motion toward sample center is hindered by the pinning centers, which tend to pin any vortex that passes by. A nonequilibrium state, the critical state, is created where the vortex density is largest in the regions where magnetic flux enters the sample. At zero temperature, the Lorentz force is balanced by the pinning force. The metastable landscape of vortices can be altered in several ways, as e.g. external current circulation or changing the applied magnetic field\cite{Kim62,Anderson62,Bean64,Brandt96,Blatter94,Brandt95}. A big effort, especially since the discovery of superconductors with higher critical temperatures, has been invested to understand the establishment of the critical state, and the nature and universality of the mechanisms behind the organization of vortices in pinning landscapes\cite{Altshuler04,Bak88,G89,Souletie,RevNagel}. In some particular situations, isolated vortices, and its motion, have been studied with great detail, at high temperatures and/or at very low magnetic fields\cite{Pardo98,Troyanovski99,Bending99,Goa01,Troyanovski02,Togawa05,Welling05,Aegerter06,Lee09}. However, there is still no direct information, at the level of single vortices, about the way the vortex system responds to alterations of the critical conditions at very low temperatures. For instance, starting with the system in a critical state, an increase of the applied magnetic field brings it to an overcritical situation, and vortices relax to a new critical state. In order to progress on the understanding of such a process, we have followed by Scanning Tunneling Microscopy and Spectroscopy, STM/S, the way that groups of around 30 vortices change their positions when an initial critical state is modified by small increments of the applied magnetic field $\delta$H$\ll$H. Experiments were performed at 100 mK to avoid thermal creep, in zero field cooled conditions and in an amorphous thin film. Vortex positions are determined by linear extended pinning centers due to small surface roughness, which can be previously imaged using the STM in topography mode. The direction of motion is conditioned by the distribution of pinning centers. Vortex bundles with roughly hexagonal vortex arrangements are gradually distorted, until they relax through small vortex avalanches.

In previous work we have shown that superconducting amorphous W-based films, fabricated using focused-ion-beam \cite{Sadki04}, can be studied with STM/S without a particular surface preparation, showing perfect BCS behavior at zero field and vortex images with many vortices in a large field and temperature range\cite{Guillamon08b,Guillamon09Nat}. Here we use the same experimental set-up. The sample is 200 nm thick, and has been grown by sweeping the focused ion beam onto a Au layer previously deposited by evaporation on a Si substrate. Vortex pinning occurs through features appearing at the surface, namely linear depressions of 5 nm height, similar to those discussed in previous work\cite{Guillamon08b,Guillamon09Nat}. Vortex lattice STS images are built from the normalized zero bias conductance changes as a function of the position in maps of 64$\times$64 points. Each image is taken at a fixed field some minutes after changing the magnetic field. Images require between 8 to 30 minutes each and are repeated several times at a given magnetic field to check for eventual vortex motion (creep) as a function of time, which was actually never observed\cite{Note1}. We have studied up to four different regions, at different magnetic fields (between 1 T and a few T), finding the behavior discussed in the following. In Fig.1(a), we show the topography of the area on which we focus here, where we have identified three main linear surface corrugation features, two in the upper left and right corners and another one in the bottom left part, marked in the figure with red dotted lines (see also \cite{EPAPS1}).

\begin{figure}[ht]
\includegraphics[clip=true,angle=0,width=9cm]{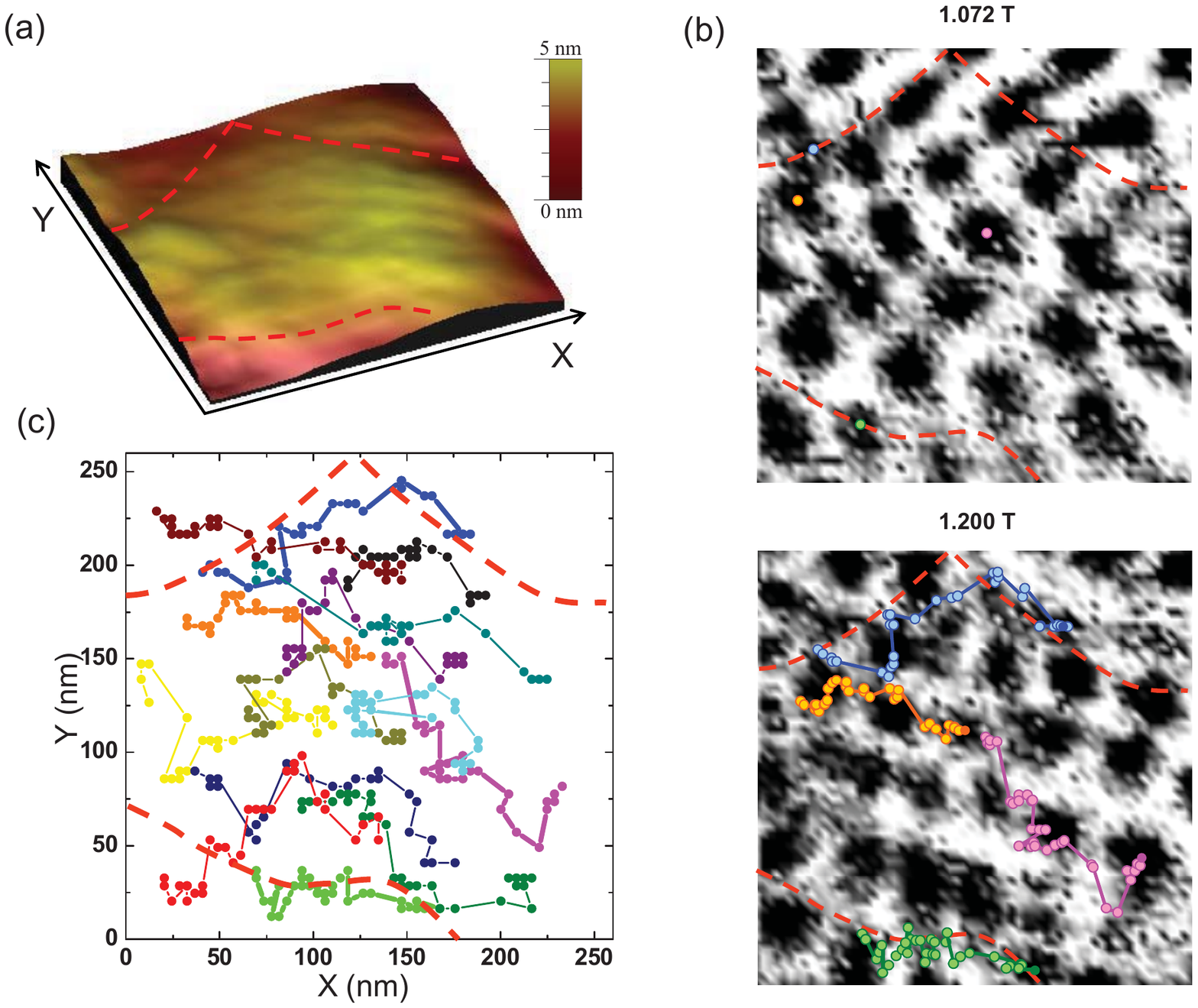}
\vskip -4,3 cm
\caption{{\small{(a)Topography (STM image) of a typical region of the W deposited film where red dotted lines mark small linear features at the surface. The lateral size of the image is of 260 nm, with a corrugation of 5 nm. Tunneling conductance is of 3$\mu$S, with 2.5 mV bias voltage. In (b) we show the initial (upper panel) and final (bottom panel) images of a series composed of 33 vortex images taken, in the area with the topography shown in (a), at 100 mK when increasing the magnetic field. The final image of the series shows the trajectories of four vortices, highlighted by colored points. (c) The trajectories of 17 vortices are represented in the x-y plane of the topography image (a). These 17 vortices stay within the scanning window during the whole sequence. Other vortices, not highlighted, jump in and out of the scanning window during the sequence. Lines joining points in (b) and (c) are guides to the eye. See Ref.\protect\cite{EPAPS1} for more details.}}\label{fig1}}
\end{figure}

In Fig. 1(b) we show superconducting vortices in the same region obtained at 100 mK, after zero field cooling and increasing H up to 1.072 T. In the upper panel of Fig. 1(b) we show the starting vortex arrangement. A vortex bundle showing distorted hexagonal order is found at the region between pinning sites. At the top and bottom part of the image, vortex arrangements are determined by pinning through the surface features. At the bottom panel of Fig. 1(b), we show the vortex image obtained after increasing the magnetic field in steps of 0.004 T up to 1.2 T\cite{EPAPS1}. Vortices of the bundle between pins (e.g. orange and magenta) tend to move to the bottom right part of the image, following a path which is roughly parallel to upper and lower linear surface depressions. On the other hand, vortices close to the linear surface pinning centers (e.g. blue and green) move along them. Note that motion occurs in smaller steps separated by large jumps, and that there is some correlation between jumps.

\begin{figure}[ht]
\includegraphics[clip=true,width=9cm,clip]{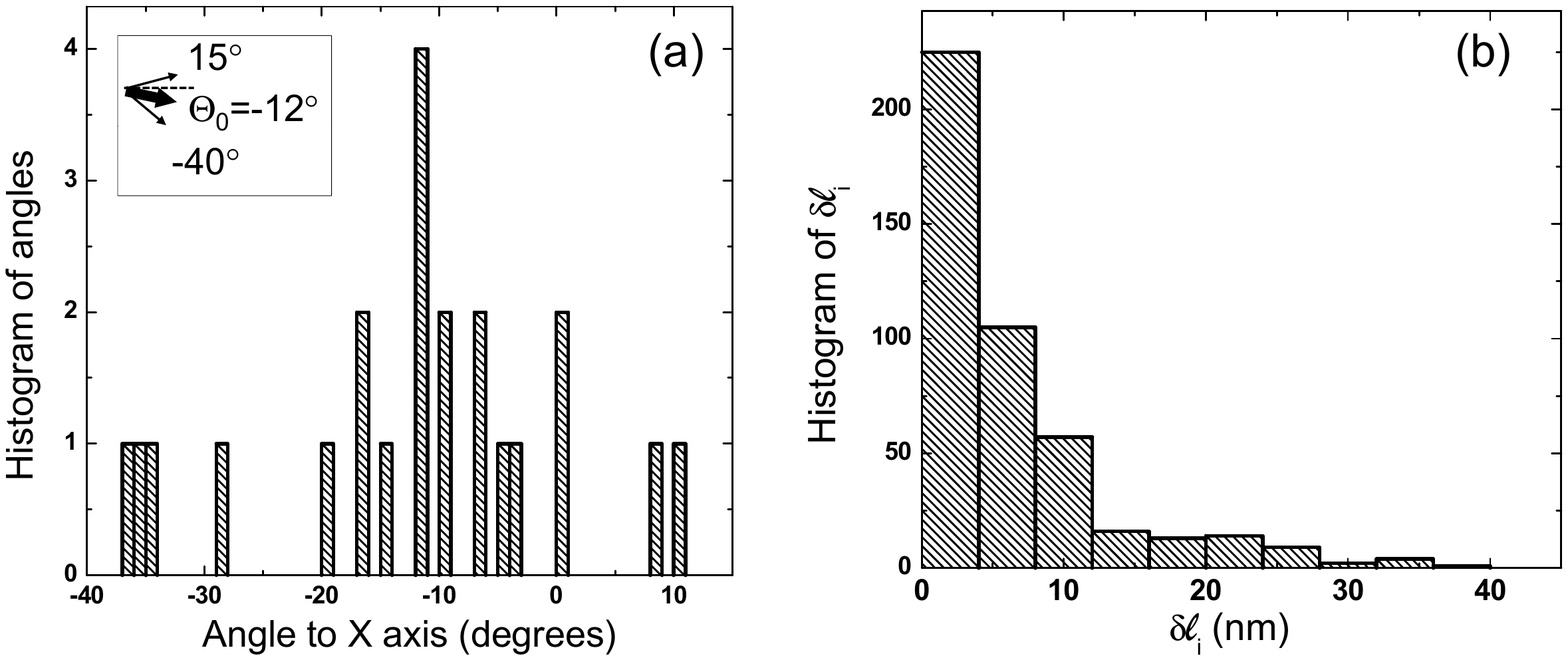}
\vskip -3cm
\caption{{\small{(a) Histogram of the angles formed by the vector joining the initial and final vortex positions and the x-axis of the image (see Fig.1). The inset shows schematically the overall vortex motion, towards an angle of around 12$^\circ$ with respect to the x-axis, and the maximum overall deviations found with respect to this direction. (b) Histogram of the size of the displacements of vortices $\delta \ell_{i}$, as defined in the text. Most frequent displacements $\delta \ell_{i}$ are of a few nm.}}\label{fig2}}
\end{figure}

All trajectories are shown in Fig. 1(c). To highlight the overall direction of vortex motion, we have plotted in Fig.2(a) a histogram of the angle $\theta$ with respect to the X axis of the vector joining initial and final positions in each vortex trajectory. There is a net flux motion flow towards a preferential direction around $\theta_{0} = -12^{\circ}$, i.e. towards the lower right part of the image at a path roughly defined by the upper and lower pinning lines, and close to a high symmetry direction of the hexagonal vortex arrangement within the bundle between pins. Clearly, a magnetic field gradient is found close to the studied region. When increasing further the magnetic field, the gradient increases and vortices move towards the direction of the maximum magnetic field gradient, being attracted to the lower right part of the image as given by $\theta_{0} = -12^{\circ}$. In Fig.2(b) we show a histogram of the size of all displacements after each magnetic field step $\delta \ell_{i}$. $\delta \ell_{i}$ is, for each vortex path, given by $\delta \ell_{i}=|(\overrightarrow{r_{i}}-\overrightarrow{r_{i-1}})|$, where $\overrightarrow{r_{i}}$ is the vortex position in magnetic field step $i$ and $\overrightarrow{r_{i-1}}$, the vortex position in previous step. Clearly, most displacements $\delta \ell_{i}$ are small, and sometimes large jumps, associated with large vortex re-organizations, occur.

\begin{figure}[ht]
\includegraphics[clip=true,width=6.5cm,angle=0,clip]{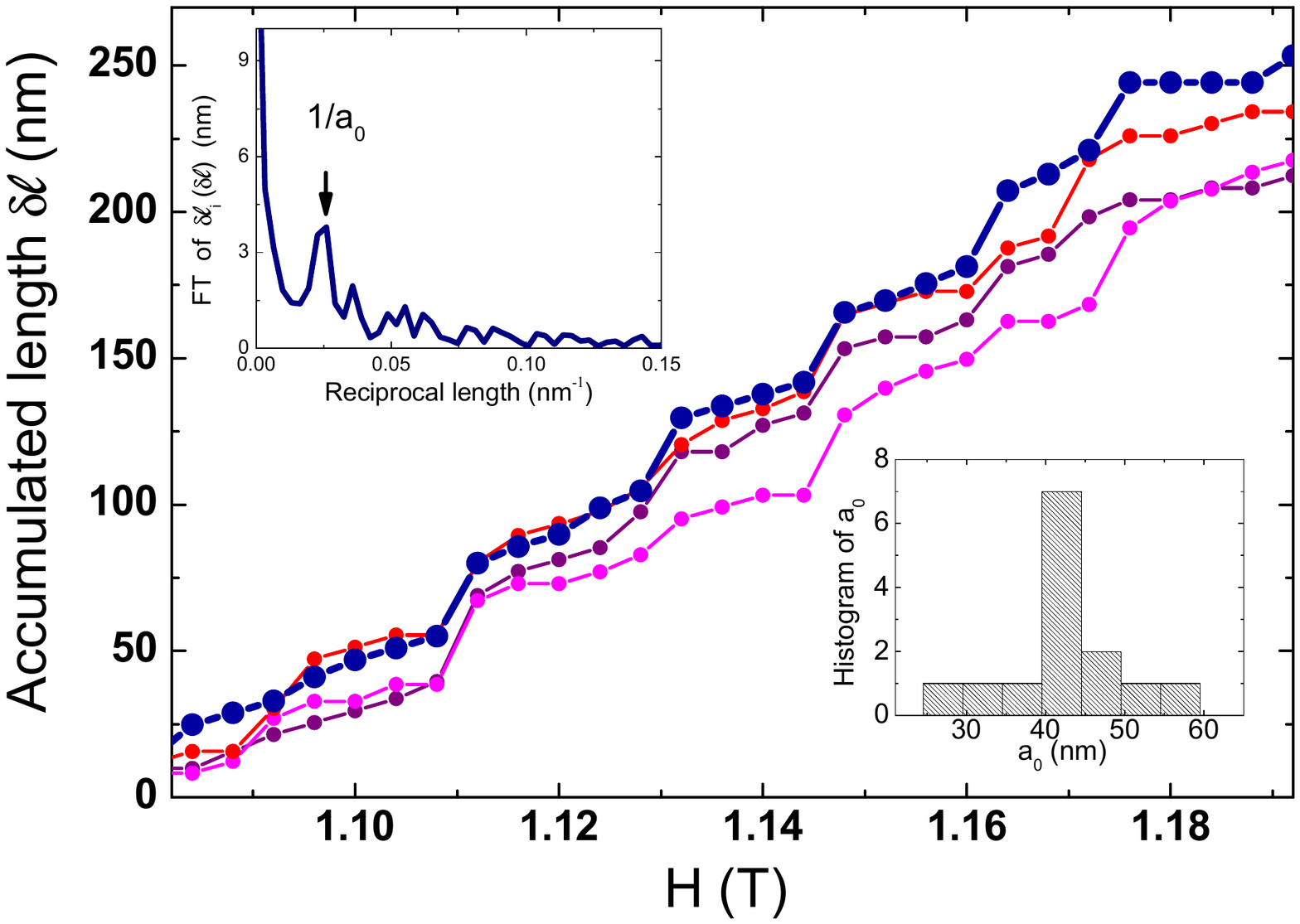}
\vskip -3,0 cm
\caption{{\small{Accumulated length $\delta l=\sum\limits_{i=1}^{i=n}{\delta l_{i}}$ (see text) vs. magnetic field of four vortex trajectories. The color code is the same as in Fig.1(c). In the upper left inset, the Fourier transform of the relative displacement $\delta l_{i}$ as a function of the accumulated length $\delta l$ for one vortex (the one with blue points in the figure) is shown. Arrow marks the peak found at the inverse of $a_0=40 nm$. The lower bottom inset shows the histogram of the inverse of the position of this peak in the Fourier transforms of the accumulated lengths of many vortex trajectories. The histogram peaks at the inverse of the expected intervortex distance $a \sim $ 43-46 nm.}}\label{fig4}}
\end{figure}

It is interesting to discuss the accumulated length $\delta \ell=\sum\limits_{i=1}^{i=n}{\delta \ell_{i}}$ of each vortex path as a function of the magnetic field. This shows the mechanism mastering the behavior of our system on its way along successive critical states. The result (Fig.3) is that there is a succession of several small size steps separated by larger jumps.

The repeated succession can be discussed by analyzing the Fourier transform of the dependence of the step size $\delta \ell_{i}$ for each vortex path as a function of $\delta \ell$. In the upper left inset of Fig.3 we show an example. There is a peak close to $a_0^{-1}$ = $(40 nm)^{-1}$. $a_0$ is close to the expected intervortex distance $a = (4/3)^{1/4}$($\Phi_{0}$/H)$^{1/2}$ for a hexagonal vortex lattice ($a$ varies from 46 nm to 43 nm between 1.072 and 1.2 T). This peak shows that the appearance of larger steps in $\delta \ell_{i}$ occurs at regular intervals of the accumulated length $\delta \ell$ of vortex paths. The peak is found close to $a$ in many vortex paths, as shown by the histogram of its position (lower right inset of Fig.3). Thus, vortex paths in our experiment are modulated with lattice periodicity. This is remarkable, as translational order disappears within a few intervortex distances. Vortex motion showing some periodicity has been observed in vortex lattices driven at high velocities and/or high temperatures under different types of pinning potentials\cite{Pardo98,Troyanovski99,Troyanovski02,Besseling05,Lee09,Harris95}. Here, however, the vortex lattice is static. The landscape of linear pinning centers in our samples allows the formation of nearly hexagonal vortex bundles in between pinning centers at some magnetic field values. The orientation and position of these nearly hexagonal vortex bundles mimics the interplay between elastic energy, tending to organize the vortices in hexagons, and pinning energy, tending to fit them to the linear features at the topography. Such hexagonal very short range order possibly creates the observed regularity in the overall accumulated vortex motion.

Further microscopic insight can be obtained by analyzing in detail vortex positions at each magnetic field step. In Fig.4 we highlight a particular interval of six steps, between 1.112 T and 1.132 T. At the beginning of the sequence, the central vortex bundle shows a nearly hexagonal arrangement. Then, during the first four small steps (from 1.112 T to 1.128 T), vortices travel small distances and the central bundle becomes gradually distorted. Vortices with five or seven nearest neighbors appear in the images, as shown in Fig.4(b) (at 1.112 T, 1.12 T and 1.128 T). In the next step, a dramatic rearrangement occurs. It involves about ten vortices, which jump at the same time over larger distances $\delta \ell_{i}$ than in previous magnetic field steps (see Fig.4(a)). These collective large jumps, small vortex avalanches, are the mechanism which governs the passage between two nearby well-differentiated vortex arrangements within a given pinning landscape. To our knowledge, these data are the first which evidence vortex avalanches at the scale of individual vortices at magnetic fields as high as a Tesla and very low temperatures. Thus, the response of our system to the quasistatic increase of the external magnetic field is composed of two different processes, distorting quasielastic small displacements, and vortex avalanches, by which all or part of the accumulated stress is relieved.

\begin{figure}[ht]
\vskip -1 cm
\includegraphics[clip=true,width=10cm,clip]{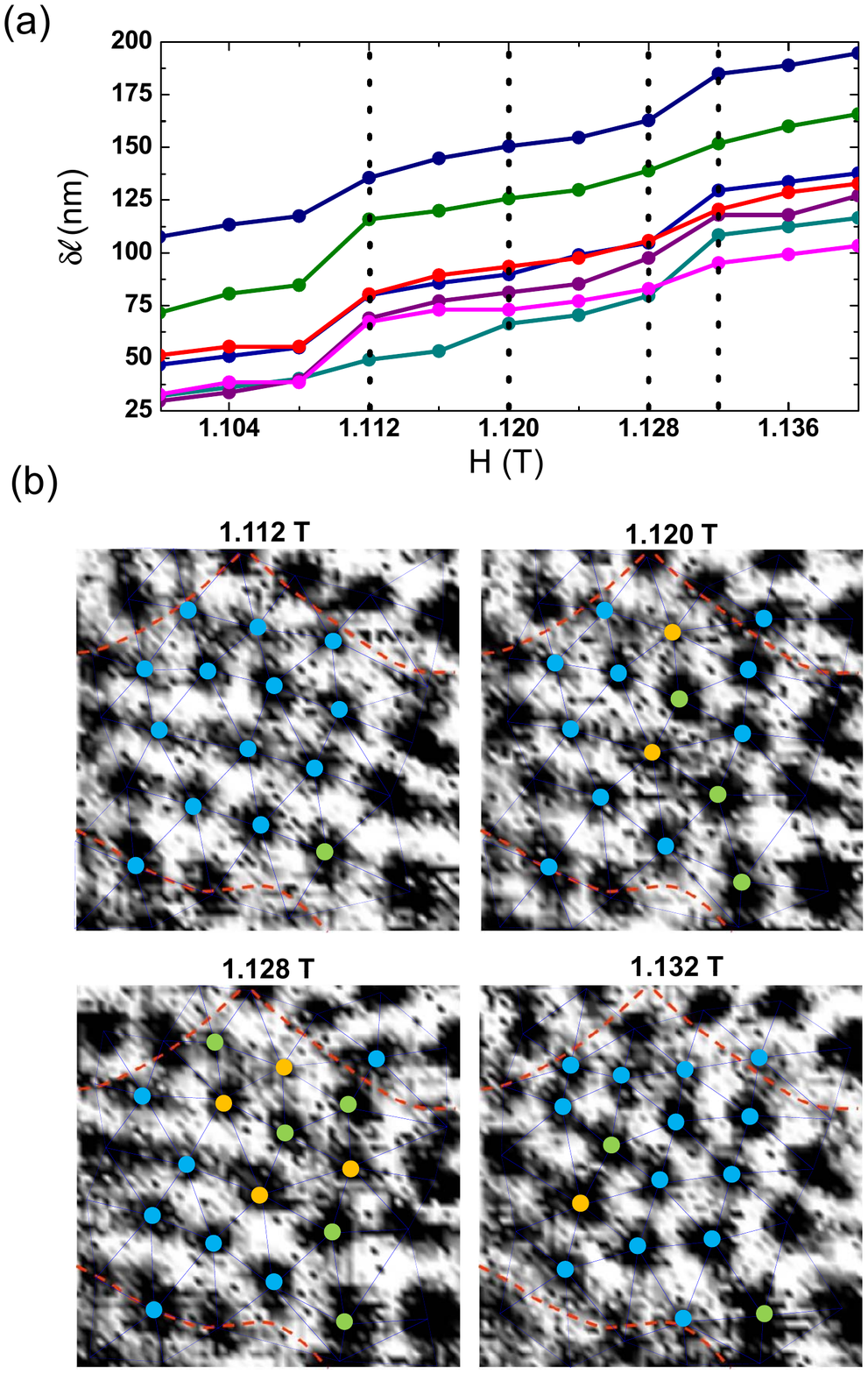}
\vskip -1 cm
\caption{{\small{In (a) we show the accumulated length $\delta l$ (see text and Fig.3) of a set of vortex trajectories in a short magnetic field interval. Color code for identifying each vortex trajectory is the same as in Figs.1 and 3. Dashed lines mark the magnetic fields where we have taken the images shown in (b). There, we show vortex arrangements at 1.112 T, 1.120 T, 1.128 T and 1.132 T. Vortices in the central bundle are marked by points and have been Delaunay triangulated. Green, blue and orange points show vortices with, respectively, five, six and seven nearest neighbors. Red dashed lines mark the pinning centers, as in Fig.1(a).}}\label{fig3}}
\end{figure}

We can make a simple estimation of the elastic properties of the bundles involved in our experiment. The pinning force F$_{p}$ can be obtained from the magnetic field, assuming that F$_{p}$ compensates the Lorentz driving force produced by the screening currents J which maintain the local magnetic field gradient\cite{Kim62,Anderson62,Bean64,Brandt96,Blatter94,Brandt95}. J = J$_{c}$, where J$_{c}$ is the critical current density, which we estimate from macroscopic measurements to be about 10$^{5}$ A/cm$^{2}$ here\cite{Sadki04,DeTeresa09}, and thus F$_{p}\approx J_{c}$B. We obtain pinning forces of around 1.1 10$^9$ N/m$^3$. The slope $\delta$F$_p$/$\delta \ell_i$ gives, within a simple Hooke's law approximation, an elastic constant. For example, we find a vortex lattice compression modulus C$_{11}$ of around 300 N/m$^2$, (taking C$_{11}=$F$_p$R$_c$a$_0$/$\delta \ell_i$ with R$_c$ being the radial correlation length\cite{Blatter94,Brandt95}) when we examine the parts of the trajectories related to the regular vortex motion previously discussed (Fig.3). The corresponding potential well is found to be around 5 K. Potential wells up to several tens of K are found, within this basic analysis, for motion involving avalanches of many vortices.

Let us note that detailed simulations of superconducting vortices within a pinning landscape given by point like pinning centers at zero temperature show a rather rich behavior, governed by vortex avalanches \cite{Reichhardt96,Olson97}. When the density of pinning sites is low, easy channels for vortex flow appear, related to interstitial vortices moving around their pinned neighbors, and at high pinning densities, vortices jump from pin to pin. In our experiment pinning centers are not point like but linear. Remarkably, vortices are able to stick to the pinning centers even if the pin is oriented perpendicular to the overall flow (as shown by the blue vortex of Fig.1 during the first part of the sequence). This occurs at the expense of allowing jump and vortex re-orientations within the central bundle.

In summary, we have studied vortex motion in the critical state of a superconductor with linear pinning centers at high magnetic fields and in the absence of thermal excitation. Our experiment gives answers some relevant questions. Perhaps the most important is about the very nature of  "avalanching objects" and has been expressed in the following way \cite{Altshuler04}: are they individual vortices, or flux bundles? are they rigid entities? Our experiments show that these objects are bundles of vortices, and that they are deformable entities. Even more, this deformability is a fundamental aspect of the way bundles move between pinning centers.

We acknowledge discussions with F. Guinea and A.I. Buzdin. The Laboratorio de Bajas Temperaturas is associated to the ICMM of the CSIC. This work was supported by the Spanish MICINN (Consolider Ingenio Molecular Nanoscience CSD2007-00010 program, MAT2008-06567-C02 and FIS2008-00454), by the Comunidad de Madrid through programs Citecnomik and Nanobiomagnet, by the Aragon Regional Governement, and by NES and ECOM programs of the ESF.


\begin{thebibliography}{30}
\expandafter\ifx\csname natexlab\endcsname\relax\def\natexlab#1{#1}\fi
\expandafter\ifx\csname bibnamefont\endcsname\relax
  \def\bibnamefont#1{#1}\fi
\expandafter\ifx\csname bibfnamefont\endcsname\relax
  \def\bibfnamefont#1{#1}\fi
\expandafter\ifx\csname citenamefont\endcsname\relax
  \def\citenamefont#1{#1}\fi
\expandafter\ifx\csname url\endcsname\relax
  \def\url#1{\texttt{#1}}\fi
\expandafter\ifx\csname urlprefix\endcsname\relax\def\urlprefix{URL }\fi
\providecommand{\bibinfo}[2]{#2}
\providecommand{\eprint}[2][]{\url{#2}}

\bibitem[{\citenamefont{Bean}(1964)}]{Bean64}
\bibinfo{author}{\bibfnamefont{C.}~\bibnamefont{Bean}}, \bibinfo{journal}{Rev.
  of Mod. Phys.} \textbf{\bibinfo{volume}{36}}, \bibinfo{pages}{31}
  (\bibinfo{year}{1964}).

\bibitem[{\citenamefont{Kim et~al.}(1962)\citenamefont{Kim, Hempstead, and
  Strnad}}]{Kim62}
\bibinfo{author}{\bibfnamefont{Y.}~\bibnamefont{Kim}},
  \bibinfo{author}{\bibfnamefont{C.}~\bibnamefont{Hempstead}},
  \bibnamefont{and} \bibinfo{author}{\bibfnamefont{A.}~\bibnamefont{Strnad}},
  \bibinfo{journal}{Phys. Rev. Lett.} \textbf{\bibinfo{volume}{9}},
  \bibinfo{pages}{306} (\bibinfo{year}{1962}).

\bibitem[{\citenamefont{Anderson}(1962)}]{Anderson62}
\bibinfo{author}{\bibfnamefont{P.}~\bibnamefont{Anderson}},
  \bibinfo{journal}{Phys. Rev. Lett.} \textbf{\bibinfo{volume}{9}},
  \bibinfo{pages}{309} (\bibinfo{year}{1962}).

\bibitem[{\citenamefont{Brandt}(1996)}]{Brandt96}
\bibinfo{author}{\bibfnamefont{E.~H.}~\bibnamefont{Brandt}},
  \bibinfo{journal}{Phys. Rev. B} \textbf{\bibinfo{volume}{54}},
  \bibinfo{pages}{4246} (\bibinfo{year}{1996}).

\bibitem[{\citenamefont{Blatter et~al.}(1994)\citenamefont{Blatter, Feigel'man,
  Geshkenbein, Larkin, and Vinokur}}]{Blatter94}
\bibinfo{author}{\bibfnamefont{G.}~\bibnamefont{Blatter}},
  \bibinfo{author}{\bibfnamefont{M.}~\bibnamefont{Feigel'man}},
  \bibinfo{author}{\bibfnamefont{V.}~\bibnamefont{Geshkenbein}},
  \bibinfo{author}{\bibfnamefont{A.}~\bibnamefont{Larkin}}, \bibnamefont{and}
  \bibinfo{author}{\bibfnamefont{V.}~\bibnamefont{Vinokur}},
  \bibinfo{journal}{Rev. Mod. Phys.} \textbf{\bibinfo{volume}{66}},
  \bibinfo{pages}{1125} (\bibinfo{year}{1994}).

\bibitem[{\citenamefont{Brandt}(1995)}]{Brandt95}
\bibinfo{author}{\bibfnamefont{E.~H.} \bibnamefont{Brandt}},
  \bibinfo{journal}{Rep. Prog. Phys.} \textbf{\bibinfo{volume}{58}},
  \bibinfo{pages}{1465} (\bibinfo{year}{1995}).

\bibitem[{\citenamefont{Altshuler and Johansen}(2004)}]{Altshuler04}
\bibinfo{author}{\bibfnamefont{E.}~\bibnamefont{Altshuler}} \bibnamefont{and}
  \bibinfo{author}{\bibfnamefont{T.}~\bibnamefont{Johansen}},
  \bibinfo{journal}{Rev. Mod. Phys.} \textbf{\bibinfo{volume}{76}},
  \bibinfo{pages}{471} (\bibinfo{year}{2004}).

\bibitem[{\citenamefont{Bak et~al.}(1988)\citenamefont{Bak, Tang, and
  Wiesenfeld}}]{Bak88}
\bibinfo{author}{\bibfnamefont{P.}~\bibnamefont{Bak}},
  \bibinfo{author}{\bibfnamefont{C.}~\bibnamefont{Tang}}, \bibnamefont{and}
  \bibinfo{author}{\bibfnamefont{K.}~\bibnamefont{Wiesenfeld}},
  \bibinfo{journal}{Phys. Rev. A} \textbf{\bibinfo{volume}{38}},
  \bibinfo{pages}{364} (\bibinfo{year}{1988}).

\bibitem[{\citenamefont{deGennes}(1989)}]{G89}
\bibinfo{author}{\bibfnamefont{P.~G.} \bibnamefont{deGennes}},
  \emph{\bibinfo{title}{Superconductivity of Metals and Alloys}}
  (\bibinfo{publisher}{Addison-Wesley}, \bibinfo{address}{New York},
  \bibinfo{year}{1989}).

\bibitem[{\citenamefont{Souletie}(1983)}]{Souletie}
\bibinfo{author}{\bibfnamefont{J.}~\bibnamefont{Souletie}},
  \bibinfo{journal}{J. Phys. (Paris)} \textbf{\bibinfo{volume}{44}},
  \bibinfo{pages}{1095} (\bibinfo{year}{1983}).

\bibitem[{\citenamefont{Nagel}(1992)}]{RevNagel}
\bibinfo{author}{\bibfnamefont{S.}~\bibnamefont{Nagel}}, \bibinfo{journal}{Rev.
  Mod. Phys.} \textbf{\bibinfo{volume}{64}}, \bibinfo{pages}{321}
  (\bibinfo{year}{1992}).

\bibitem[{\citenamefont{Pardo et~al.}(1998)\citenamefont{Pardo, de~la Cruz,
  Gammel, Bucher, and Bishop}}]{Pardo98}
\bibinfo{author}{\bibfnamefont{F.}~\bibnamefont{Pardo}},
  \bibinfo{author}{\bibfnamefont{F.}~\bibnamefont{de~la Cruz}},
  \bibinfo{author}{\bibfnamefont{P.}~\bibnamefont{Gammel}},
  \bibinfo{author}{\bibfnamefont{E.}~\bibnamefont{Bucher}}, \bibnamefont{and}
  \bibinfo{author}{\bibfnamefont{D.}~\bibnamefont{Bishop}},
  \bibinfo{journal}{Nature} \textbf{\bibinfo{volume}{396}},
  \bibinfo{pages}{348} (\bibinfo{year}{1998}).

\bibitem[{\citenamefont{Troyanovski et~al.}(1999)\citenamefont{Troyanovski,
  Aarts, and Kes}}]{Troyanovski99}
\bibinfo{author}{\bibfnamefont{A.~M.} \bibnamefont{Troyanovski}},
  \bibinfo{author}{\bibfnamefont{J.}~\bibnamefont{Aarts}}, \bibnamefont{and}
  \bibinfo{author}{\bibfnamefont{P.~H.} \bibnamefont{Kes}},
  \bibinfo{journal}{Nature} \textbf{\bibinfo{volume}{399}},
  \bibinfo{pages}{665} (\bibinfo{year}{1999}).

\bibitem[{\citenamefont{Bending}(1999)}]{Bending99}
\bibinfo{author}{\bibfnamefont{S.}~\bibnamefont{Bending}},
  \bibinfo{journal}{Adv. Phys.} \textbf{\bibinfo{volume}{48}},
  \bibinfo{pages}{449} (\bibinfo{year}{1999}).

\bibitem[{\citenamefont{Goa et~al.}(2001)\citenamefont{Goa, Haughlin,
  Baziljevich, Il'yashenko, Gammel, and Johansen}}]{Goa01}
\bibinfo{author}{\bibfnamefont{P.}~\bibnamefont{Goa}},
  \bibinfo{author}{\bibfnamefont{et}~\bibnamefont{al.}},
  \bibinfo{journal}{Superc. Sci. Technol.} \textbf{\bibinfo{volume}{14}},
  \bibinfo{pages}{729} (\bibinfo{year}{2001}).

\bibitem[{\citenamefont{Troyanovski et~al.}(2002)\citenamefont{Troyanovski, van
  Hecke, Saha, Aarts, and Kes}}]{Troyanovski02}
\bibinfo{author}{\bibfnamefont{A.~M.} \bibnamefont{Troyanovski}},
  \bibinfo{author}{\bibfnamefont{M.}~\bibnamefont{van Hecke}},
  \bibinfo{author}{\bibfnamefont{N.}~\bibnamefont{Saha}},
  \bibinfo{author}{\bibfnamefont{J.}~\bibnamefont{Aarts}}, \bibnamefont{and}
  \bibinfo{author}{\bibfnamefont{P.~H.} \bibnamefont{Kes}},
  \bibinfo{journal}{Phys. Rev. Lett.} \textbf{\bibinfo{volume}{89}},
  \bibinfo{pages}{147006} (\bibinfo{year}{2002}).

\bibitem[{\citenamefont{Togawa et~al.}(2005)\citenamefont{Togawa, Harada,
  Akashi, Kasai, Matsuda, Nori, Maeda, and Tonomura}}]{Togawa05}
\bibinfo{author}{\bibfnamefont{Y.}~\bibnamefont{Togawa}},
  \bibinfo{author}{\bibfnamefont{et}~\bibnamefont{al.}},
  \bibinfo{journal}{Phys. Rev. Lett.} \textbf{\bibinfo{volume}{95}},
  \bibinfo{pages}{087002} (\bibinfo{year}{2005}).

\bibitem[{\citenamefont{Welling et~al.}(2005)\citenamefont{Welling, Aegerter,
  and Wijngaarden}}]{Welling05}
\bibinfo{author}{\bibfnamefont{M.~S.}~\bibnamefont{Welling}},
  \bibinfo{author}{\bibfnamefont{C.~M.}~\bibnamefont{Aegerter}}, \bibnamefont{and}
  \bibinfo{author}{\bibfnamefont{R.~J.}~\bibnamefont{Wijngaarden}},
  \bibinfo{journal}{Phys. Rev. B} \textbf{\bibinfo{volume}{71}},
  \bibinfo{pages}{104515} (\bibinfo{year}{2005}).

\bibitem[{\citenamefont{Aegerter et~al.}(2006)\citenamefont{Aegerter, Welling,
  and Wijngaarden}}]{Aegerter06}
\bibinfo{author}{\bibfnamefont{C.}~\bibnamefont{Aegerter}},
  \bibinfo{author}{\bibfnamefont{M.}~\bibnamefont{Welling}}, \bibnamefont{and}
  \bibinfo{author}{\bibfnamefont{R.}~\bibnamefont{Wijngaarden}},
  \bibinfo{journal}{Europhys. Lett.} \textbf{\bibinfo{volume}{74}},
  \bibinfo{pages}{397} (\bibinfo{year}{2006}).

\bibitem[{\citenamefont{Lee et~al.}(2009)\citenamefont{Lee, Wang, Dreyer,
  Berger, and Barker}}]{Lee09}
\bibinfo{author}{\bibfnamefont{J.}~\bibnamefont{Lee}},
  \bibinfo{author}{\bibfnamefont{H.}~\bibnamefont{Wang}},
  \bibinfo{author}{\bibfnamefont{M.}~\bibnamefont{Dreyer}},
  \bibinfo{author}{\bibfnamefont{H.}~\bibnamefont{Berger}}, \bibnamefont{and}
  \bibinfo{author}{\bibfnamefont{B.}~\bibnamefont{Barker}},
  \bibinfo{journal}{cond-mat} p. \bibinfo{pages}{0902.0452v1}
  (\bibinfo{year}{2009}).

\bibitem[{\citenamefont{Sadki et~al.}(2004)\citenamefont{Sadki, Ooi, and
  Hirata}}]{Sadki04}
\bibinfo{author}{\bibfnamefont{E.}~\bibnamefont{Sadki}},
  \bibinfo{author}{\bibfnamefont{S.}~\bibnamefont{Ooi}}, \bibnamefont{and}
  \bibinfo{author}{\bibfnamefont{K.}~\bibnamefont{Hirata}},
  \bibinfo{journal}{Appl. Phys. Lett.} \textbf{\bibinfo{volume}{85}},
  \bibinfo{pages}{6206} (\bibinfo{year}{2004}).

\bibitem[{\citenamefont{Guillamon et~al.}(2008)\citenamefont{Guillamon,
  Suderow, Vieira, Fernandez-Pacheco, Sese, Cordoba, Teresa, and
  Ibarra}}]{Guillamon08b}
\bibinfo{author}{\bibfnamefont{I.}~\bibnamefont{Guillamon}},
  \bibinfo{author}{\bibfnamefont{et}~\bibnamefont{al.}},
  \bibinfo{journal}{New Journal of Physics} \textbf{\bibinfo{volume}{10}},
  \bibinfo{pages}{093005} (\bibinfo{year}{2008}).

\bibitem[{\citenamefont{Guillamon et~al.}(2009)\citenamefont{Guillamon,
  Suderow, Fernandez-Pacheco, Sese, Cordoba, Teresa, Ibarra, and
  Vieira}}]{Guillamon09Nat}
\bibinfo{author}{\bibfnamefont{I.}~\bibnamefont{Guillamon}},
  \bibinfo{author}{\bibfnamefont{et}~\bibnamefont{al.}},
  \bibinfo{journal}{Nature Physics} \textbf{\bibinfo{volume}{5}},
  \bibinfo{pages}{651} (\bibinfo{year}{2009}).

\bibitem[{EPA()}]{Note1}
\bibinfo{note}{Dislocation motion in ordered lattices has been studied recently by K. Uchiyama et al. Physica C, in press (2010).}

\bibitem[{EPA()}]{EPAPS1}
\bibinfo{note}{See EPAPS Document No. [] for all vortex images. Each vortex is identified by colored circles with numbers. 17 vortices remain in the imaging window, whereas the rest enters the window or leaves it when changing the magnetic field. A line scan over the topography of the sample through one linear surface depression is also given. See EPAPS Document No. []  for a sequence showing consecutively all images. For more information on EPAPS, see http://www.aip.org/pubservs/epaps.html.}

\bibitem[{\citenamefont{Bessenling et~al.}(2005)\citenamefont{Bessenling, Kes,
  Drose, and Vinokur}}]{Besseling05}
\bibinfo{author}{\bibfnamefont{R.}~\bibnamefont{Bessenling}},
  \bibinfo{author}{\bibfnamefont{P.}~\bibnamefont{Kes}},
  \bibinfo{author}{\bibfnamefont{T.}~\bibnamefont{Drose}}, \bibnamefont{and}
  \bibinfo{author}{\bibfnamefont{V.}~\bibnamefont{Vinokur}},
  \bibinfo{journal}{New Journal of Physics} \textbf{\bibinfo{volume}{7}},
  \bibinfo{pages}{71} (\bibinfo{year}{2005}).

\bibitem[{\citenamefont{Harris et~al.}(1995)\citenamefont{Harris, Ong, Gagnon,
  and Taillefer}}]{Harris95}
\bibinfo{author}{\bibfnamefont{J.~M.}~\bibnamefont{Harris}},
  \bibinfo{author}{\bibfnamefont{N.~P.}~\bibnamefont{Ong}},
  \bibinfo{author}{\bibfnamefont{R.}~\bibnamefont{Gagnon}}, \bibnamefont{and}
  \bibinfo{author}{\bibfnamefont{L.}~\bibnamefont{Taillefer}},
  \bibinfo{journal}{Phys. Rev. Lett.} \textbf{\bibinfo{volume}{74}},
  \bibinfo{pages}{3684} (\bibinfo{year}{1995}).

\bibitem[{\citenamefont{Teresa et~al.}(2009)\citenamefont{Teresa,
  Fernandez-Pacheco, Cordoba, Sese, Ibarra, Guillamon, Suderow, and
  Vieira}}]{DeTeresa09}
\bibinfo{author}{\bibfnamefont{J.~M.~De} \bibnamefont{Teresa}},
  \bibinfo{author}{\bibfnamefont{et}~\bibnamefont{al.}},
  \bibinfo{journal}{MRS Proceedings} \textbf{\bibinfo{volume}{CC04-09}},
  \bibinfo{pages}{1180} (\bibinfo{year}{2009}).

\bibitem[{\citenamefont{Reichhardt et~al.}(1996)\citenamefont{Reichhardt,
  Olson, Groth, Field, and Nori}}]{Reichhardt96}
\bibinfo{author}{\bibfnamefont{C.}~\bibnamefont{Reichhardt}},
  \bibinfo{author}{\bibfnamefont{C.~J.}~\bibnamefont{Olson}},
  \bibinfo{author}{\bibfnamefont{J.}~\bibnamefont{Groth}},
  \bibinfo{author}{\bibfnamefont{S.}~\bibnamefont{Field}}, \bibnamefont{and}
  \bibinfo{author}{\bibfnamefont{F.}~\bibnamefont{Nori}},
  \bibinfo{journal}{Phys. Rev. B} \textbf{\bibinfo{volume}{53}},
  \bibinfo{pages}{R8898} (\bibinfo{year}{1996}).

\bibitem[{\citenamefont{Olson et~al.}(1997)\citenamefont{Olson, Reichhardt, and
  Nori}}]{Olson97}
\bibinfo{author}{\bibfnamefont{C.~J.}~\bibnamefont{Olson}},
  \bibinfo{author}{\bibfnamefont{C.}~\bibnamefont{Reichhardt}},
  \bibnamefont{and} \bibinfo{author}{\bibfnamefont{F.}~\bibnamefont{Nori}},
  \bibinfo{journal}{Phys. Rev. B} \textbf{\bibinfo{volume}{56}},
  \bibinfo{pages}{6175} (\bibinfo{year}{1997}).

\end{thebibliography}

\newpage

\begin{figure}[ht]
\vskip -1 cm
\includegraphics[clip=true,width=20cm,clip]{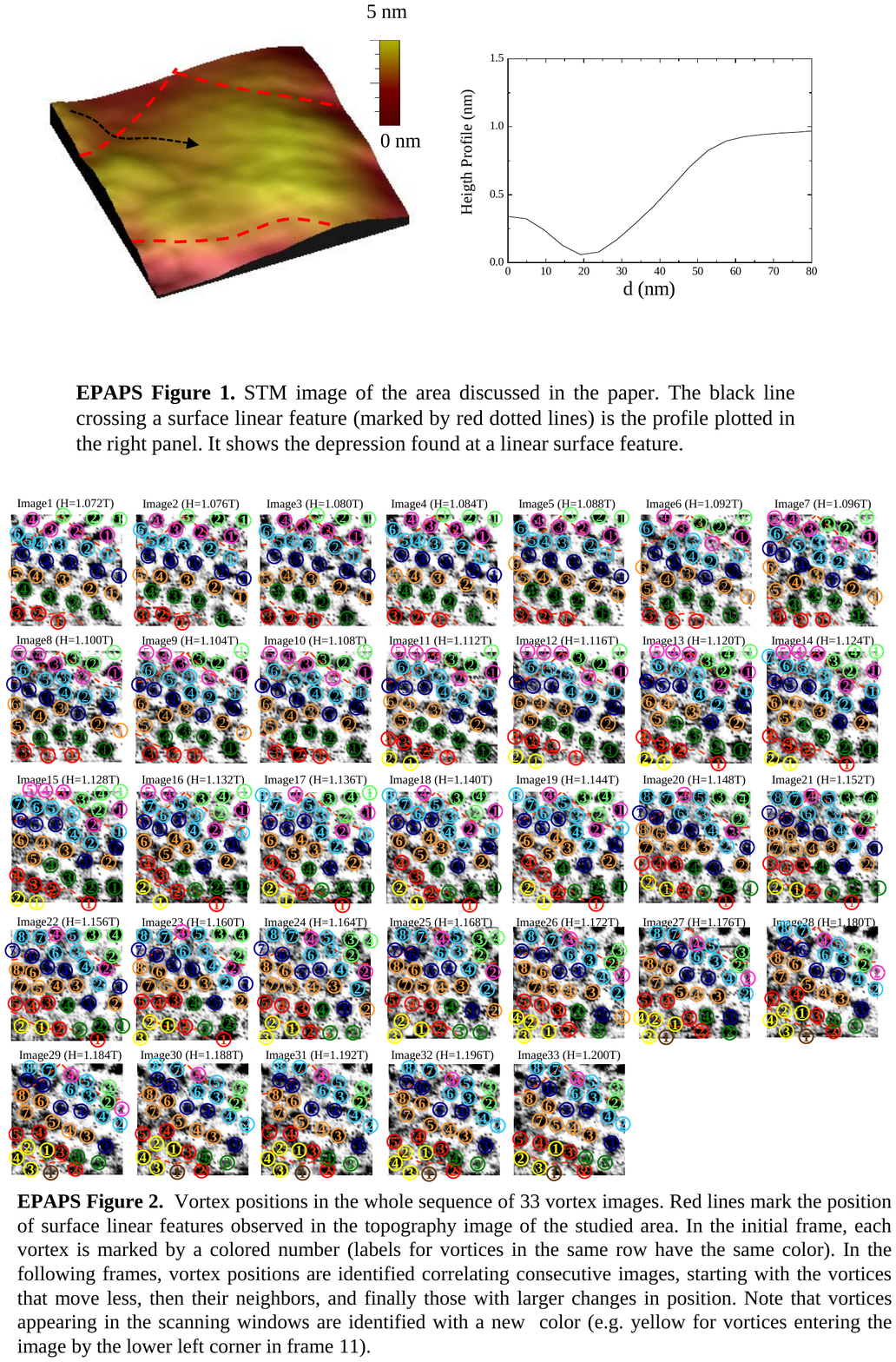}
\caption{Additional EPAPS figures. To download the video corresponding to the sequence see EPAPS server.}
\end{figure}

\end{document}